\documentclass[aps,prb,twocolumn,showpacs,floatfix,superscriptaddress,byrevtex]{revtex4-2}
\pdfoutput=1
\usepackage{soul}
\usepackage{amsmath}
\usepackage{amssymb}
\usepackage{amstext}
\usepackage{amsopn}
\usepackage{amsfonts}
\usepackage{amsxtra}
\usepackage{dcolumn}
\usepackage{graphicx}
\usepackage{bm}
\usepackage{hyperref}
\usepackage{xcolor}
\definecolor{linkcol}{rgb}{0,0,0.4} 
\definecolor{violet}{rgb}{0.32,0,0.52} 
\definecolor{citecol}{rgb}{0.5,0,0} 
\usepackage[expansion=true]{microtype}
\begin{document}
\title{Pressure-induced structural transitions triggering dimensional crossover in lithium purple bronze Li$_{0.9}$Mo$_6$O$_{17}$}
%
\author{M. K. Tran}
\affiliation{D\'{e}partement de Physique de la Mati\`{e}re Quantique, Universit\'{e} de Gen\`{e}ve, Quai Ernest-Ansermet 24, CH-1211 Gen\`{e}ve 4, Switzerland}
\author{A. Akrap}
\affiliation{D\'{e}partement de Physique de la Mati\`{e}re Quantique, Universit\'{e} de Gen\`{e}ve, Quai Ernest-Ansermet 24, CH-1211 Gen\`{e}ve 4, Switzerland}
\affiliation{D\'{e}partement de Physique, Universit\'{e} de Fribourg, Chemin du Mus\'ee, CH-1700 Fribourg, Switzerland}
\author{J. Levallois}
\affiliation{D\'{e}partement de Physique de la Mati\`{e}re Quantique, Universit\'{e} de Gen\`{e}ve, Quai Ernest-Ansermet 24, CH-1211 Gen\`{e}ve 4, Switzerland}
\author{J. Teyssier}
\affiliation{D\'{e}partement de Physique de la Mati\`{e}re Quantique, Universit\'{e} de Gen\`{e}ve, Quai Ernest-Ansermet 24, CH-1211 Gen\`{e}ve 4, Switzerland}
\author{P. Schouwink}
\affiliation{D\'{e}partement de Physique de la Mati\`{e}re Quantique, Universit\'{e} de Gen\`{e}ve, Quai Ernest-Ansermet 24, CH-1211 Gen\`{e}ve 4, Switzerland}
\affiliation{Institut des sciences et ing\'{e}nierie chimiques, EPFL, Lausanne, Switzerland}
\author{C. Besnard}
\affiliation{D\'{e}partement de Physique de la Mati\`{e}re Quantique, Universit\'{e} de Gen\`{e}ve, Quai Ernest-Ansermet 24, CH-1211 Gen\`{e}ve 4, Switzerland}
\author{P.~Lerch}
\affiliation{Infrared Beamline, Swiss Light Source, Paul Scherrer Institute, CH-5232 Villigen-PSI,~Switzerland} 
\author{J. W. Allen}
\affiliation{Randall Laboratory, University of Michigan, Ann Arbor, MI 48109, USA}
\author{M. Greenblatt}
\affiliation{Department of Chemistry and Chemical Biology, Rutgers University, 123 Bevier Road, Piscataway, NJ 08854}
\author{D. van der Marel}
\affiliation{D\'{e}partement de Physique de la Mati\`{e}re Quantique, Universit\'{e} de Gen\`{e}ve, Quai Ernest-Ansermet 24, CH-1211 Gen\`{e}ve 4, Switzerland}
%
%
\date{\today}
\begin{abstract}
At ambient pressure, lithium molybdenum purple bronze (Li$_{0.9}$Mo$_6$O$_{17}$) is a quasi-one dimensional solid in which the anisotropic crystal structure and the linear dispersion of the underlying bands produced by electronic correlations possibly bring about a rare experimental realization of Tomomaga-Luttinger liquid physics. It is also the sole member of the broader purple molybdenum bronzes family where a Peierls instability has not been identified at low temperatures. The present study reports a pressure-induced series of phase transitions between 0 and 12~GPa. These transitions are strongly reflected in infrared spectroscopy, Raman spectroscopy, and x-ray diffraction. The most dramatic effect seen in optical conductivity is the metalization of the $c$-axis, concomitant to the decrease of conductivity along the $b$-axis. 
This indicates that high pressure drives the material away from its quasi-one dimensional behavior at ambient pressure. 
While the first pressure-induced structure of the series is resolved, the identification of the underlying mechanisms driving the dimensional change in the physics remains a challenge.
\end{abstract}
\maketitle

\section{Introduction}
The family of purple bronzes gives access to a broad array of ground states: amongst them superconductor, charge or spin density wave, metal-insulator transition, or a possible realization of Tomonaga Luttinger liquid.
The purple molybdenum bronze family $\text{A(Mo}_6\text{O}_{17})$, $\text{A}=\text{Tl}, \text{Li}, \text{Na}$ and K was intensively investigated since the 1960s and each of its members was found to be a charge density wave material; the exception being the lithium molybdenum bronze~\cite{greenblatt1988}.
This system provides a very interesting experimental playground to investigate the signature of strongly correlated one dimensional physics, i.e. Tomonaga-Luttinger liquids, present in particular in the spectral function of ARPES studies~\cite{denlinger1999,wang2006}.

The transport properties of Li$_{0.9}$Mo$_6$O$_{17}$ are very anisotropic~\cite{greenblatt1984,mercure2012}, as three orders of magnitude stand between the most insulating axis ($a$) and the most conducting one ($b$). The $b$ axis resistivity displays a metallic behavior down to $\sim 25~$K where a sharp upturn occurs which is eventually followed in some samples by a superconducting transition below 2~K.
The most debated aspect of Li$_{0.9}$Mo$_6$O$_{17}$ is the cause of the 25~K resistivity upturn~\cite{dudy2012,dudy2018}. The low dimensional behavior of the material and its Fermi surface prone to nesting have, from the early studies, suggested the opening of a charge density wave (CDW) gap as the explanation for the resistivity upturn~\cite{greenblatt1984}.  
However, no evidence for a CDW has been reported using structural probes such as x-ray and neutron diffraction, and recent observations of  the charge state using NMR show that the electric field gradient and the distribution remain constant as a function of temperature, indicating the absence of a CDW~\cite{wu2019a,wu2019b}.

\begin{figure*}
\centering
\includegraphics[width=\textwidth]{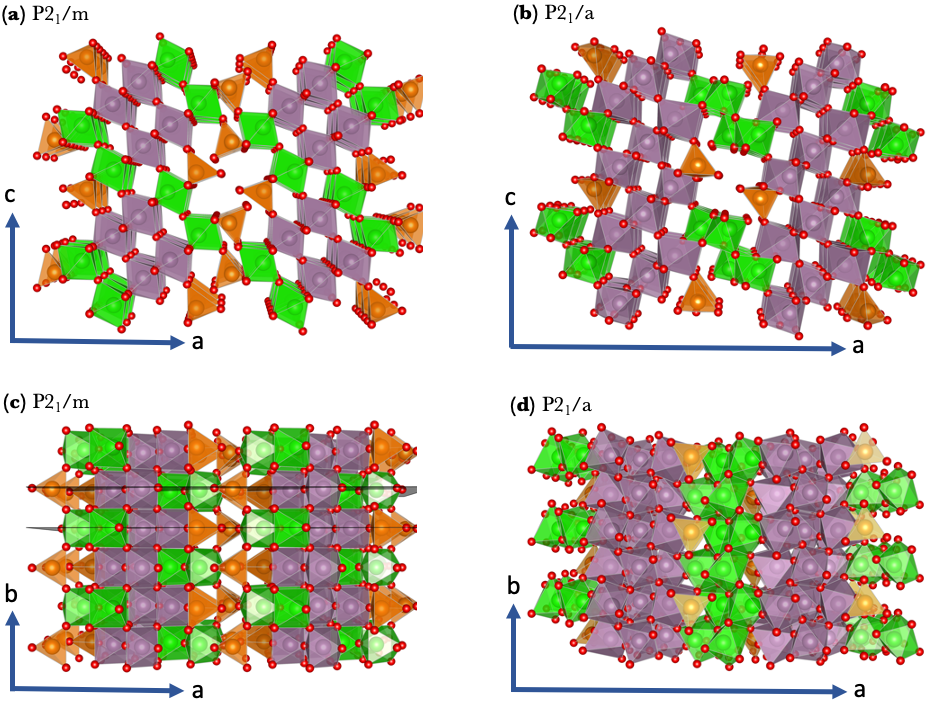}
\caption{Crystal structure at low pressure (panels {\bf a} and {\bf c}) and for pressure between 3.6 and 6~GPa (panels {\bf b} and {\bf d}). To avoid clutter the Li atoms are not shown. The projection is along the $b$ ($c$) axis for the panels {\bf a} and {\bf b} ({\bf c} and {\bf d}).
The lattice parameters are given in Table~\ref{table1}.
The horizontal planes in panel {\bf c} are mirror planes. 
This graph was generated with the open source program Vesta~\cite{vesta} using the structural parameters available in Ref.~\onlinecite{yareta2021}.} 
\label{figure1}
\end{figure*}
Li$_{0.9}$Mo$_6$O$_{17}$  was initially classified as a quasi-two-dimensional material with strong transport anisotropy, but following the resolution of its crystal structure~\cite{onoda1987} it was considered as a quasi-one-dimensional material. 
The structure, shown in Fig.~\ref{figure1}, is described by the monoclinic $P2_1/m$ space group. 
The main structural elements are MoO$_6$ octahedra organized in slabs consisting of 4 layers of octahedra terminated by MO$_4$ tetrahedra on either side. The Li ions (not shown in  Fig.~\ref{figure1}) are situated between the slabs. In this paper we will follow the convention used by Onada {\em et al.}~\cite{onoda1987} where the unique axis, $b$, and the $c$-axis at right angles with $b$, are both parallel to the slabs. The $a$-axis has an angle $\beta=90.61^\circ$ with $c$ and it is perpendicular to $b$. The lattice parameters are given in Table~\ref{table1}.
A different (but equally valid) axis labeling convention that interchanges $a$ and $c$ is used in much of the literature, especially papers reporting experimental results, so it is important to check which convention is used in any particular paper.
A pair of octahedra, individually made out of one molybdenum atom surrounded by six oxygen atoms, forms two well-separated zig-zag chains running along the $b$ axis. 
The anisotropy in the transport properties identifies the $b$-axis as the most conducting direction, $a$ as the most insulating one and the $c$-axis in between these two (see {\em e.g.} Ref.~\onlinecite{mercure2012} but pay attention to the reversed choice of $a$-~and $c$-axis) suggesting that conduction takes place primarily within the MoO$_6$ octahedra zig-zag chains.
\begin{center}
\begin{table*}[!!ht!!]
\begin{tabular}{|c|c|c|c|c|c|c|c|c|c|}
\hline 
Space group&a&b&c&$\beta$&$N_{fu/pc}$&$V_{fu}$ &Pressure&Probe\\
\hline
&\AA&\AA&\AA&degrees&-&\AA$^3$&Gpa&-\\
\hline
$P2_1/m$&12.451(9)&5.486(3)&9.347(7)&90.7(2)&2&319&$0-3.6$&XRD,IR,R\\
$P2_1/m\hspace{1mm}\&\hspace{1mm}P2_1/a$&-&-&-&-&-&-&3-3.6&XRD,IR,R\\
$P2_1/a$&23.587(8)&5.11(1)&9.495(6)&93.3(1)&4&285&$3-6$&XRD,IR,R\\
unresolved&11.34(4)&4.94(3)&9.42(2)&97.4(2)&2&261&$6-9$&XRD,IR,R\\
-&-&-&-&-&-&-&$9-11$ &IR,R\\
-&-&-&-&-&-&-&$>11$ &R\\
\hline
\end{tabular}
\caption{Space group and lattice parameters from x-ray diffraction for the different pressure ranges of the crystallographic phases of purple bronze. 
For the low-pressure phase (space group $P2_1/m$) we adopt the convention used by Onada {\em et al.}~\cite{onoda1987} where $b$ and $c$ are parallel to the slabs. To facilitate comparison of the crystal axis of the different phases we use $P2_1/a$ for the high-pressure phase instead of the standard setting $P2_1/c$.  
$N_{fu/pc}$ represents the number of formula units per primitive cell and $V_{fu}$ the volume per formula unit.  
Between 3 and 3.6 GPa $P2_1/m$ and $P2_1/a$ are observed together. 
The last column indicates the experimental probe with which the phase was observed. XRD, IR, R refer to x-ray diffraction, infrared spectroscopy and Raman spectroscopy respectively. 
\label{table1}}       
\end{table*}
\end{center}

A recent theoretical study~\cite{dudy2018} employing the Nth-order muffin-tin orbital (NMTO) method has obtained an analytical representation of the six lowest energy Mo $t_{2g}$ bands, which are found to be essentially two dimensional.  Strong covalent bonding fully gaps four of these bands, leaving in the gap two nearly degenerate metallic chain bands crossing $E_F$ to define the Fermi surface.  The dispersion of these two chain bands is highly one-dimensional, very strong along $k_b$, two orders of magnitude weaker along $k_c$ and essentially negligible along $k_a$ (same axis definitions used as in the present paper), in good qualitative agreement with the observed resistivity anisotropy.  The simplest model conceptualization is that of conducting chains with a weak inter-chain hopping parameter $t_{\perp}$~\cite{boies1995,giamarchi2009} as might be imagined from Fig.~\ref{figure1} or from the first figure of the work of Lu {\em et al.}~\cite{lu2019} (but taking note of their reversed choice of $a$ and $c$ axes).  However, in the detailed study of Ref. ~\onlinecite{dudy2018}, the effective downfolded perpendicular couplings in Li$_{0.9}$Mo$_6$O$_{17}$ are found to be of much longer range and cannot be modeled in a simple tight binding fashion, due to a resonant energy dependence arising from mixing with the four gapped bands.  These couplings lead to a splitting and a complex warping and shaping of the two-band Fermi surface, in a way that is very sensitive to the value of $E_F$, {\em i.e.} to the stoichiometry of the sample. 

Previous optical investigations have reported a very anisotropic response~\cite{degiorgi1988,choi2004}. A detailed study of the 
infrared properties~\cite{choi2004} showed free carrier optical conductivity along the (most conducting) $b$-axis of order 2 kS/cm ---a value in line with bad metallicity of Li$_{0.9}$Mo$_6$O$_{17}$--- and two orders of magnitude smaller conductivity in the direction perpendicular to $b$ in the cleavage plane (the $c$-axis, in the notation of the present paper).
Moreover, for all temperatures the low frequency response strongly differs from Drude behavior. In particular the conductivity is almost frequency independent at low frequencies, and exhibits a strong peak at 70 meV, as well as smaller peaks at lower energy, whose intensities 
depend strongly on temperature~\cite{choi2004}. The non-Drude response may be caused by low dimensionality in combination with charge carrier localization, electron correlation, or dimensional crossover~\cite{chudzinski2012}.

\begin{figure}[!!b]
\centering
\includegraphics[width=\columnwidth]{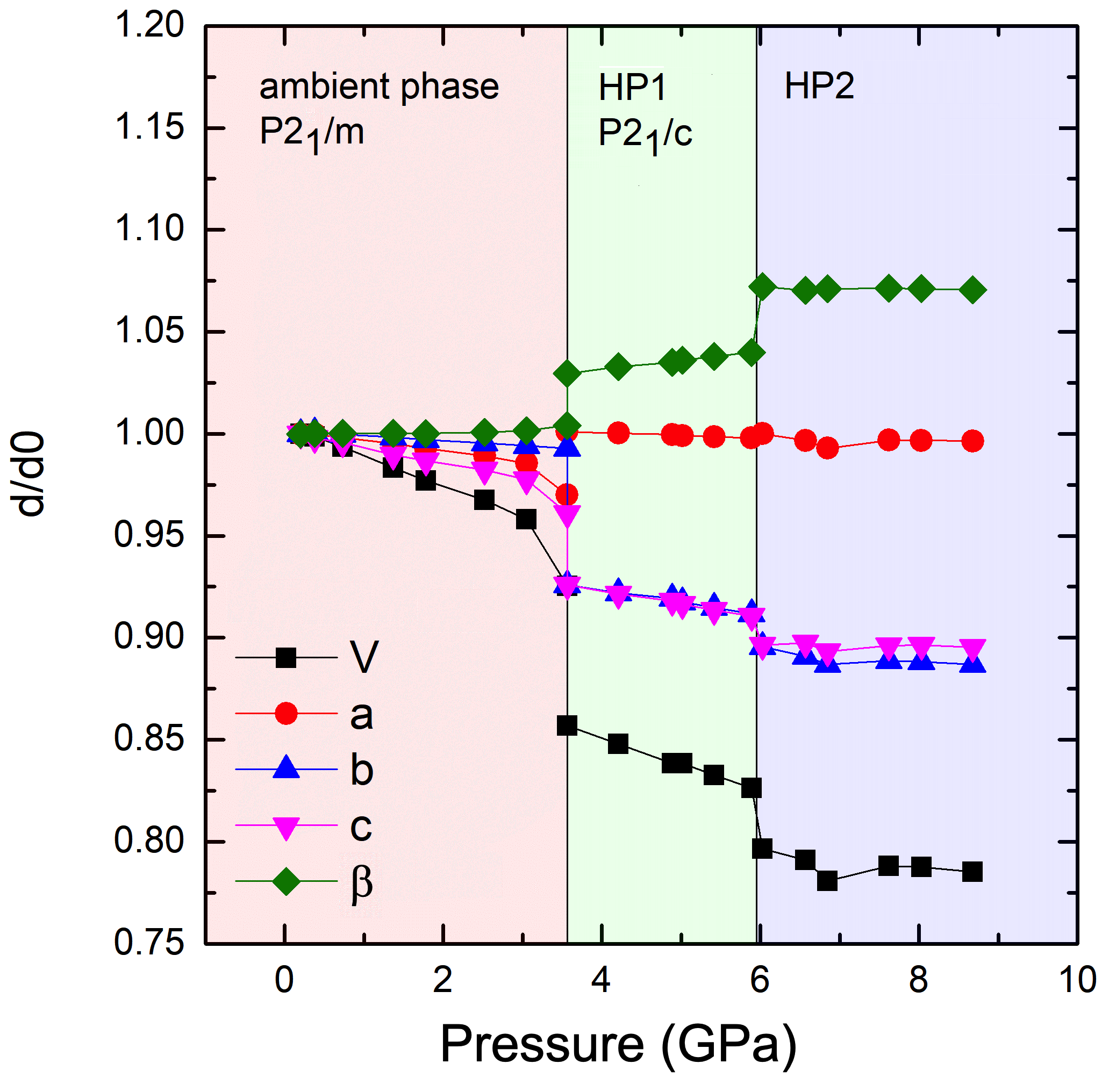}
\caption{Pressure-dependence of the lattice parameters $d$ ($d = a, b, c,\beta, V$) relative to the value at zero pressure ($d_0$), determined by Le Bail refinements using powder diffraction data. For the high pressure phase 2 (HP2) the space group could not be resolved.} 
\label{figure2}
\end{figure}
Electronic transport under pressure showed a stabilization of superconductivity at the expense of the resistivity upturn~\cite{filippini1989}. As the application of pressure generally tends to enhance the orbital overlaps in a low-dimensional solid, other ground states might be favored~\cite{sipos2008}. From the theoretical point of view, a crossover from a Luttinger liquid behavior to a 2D metallic behavior was proposed~\cite{berthod2006,chudzinski2012} for 1D conducting chains with the introduction of a small coupling. Reports of such a crossover are supported by thermal expansion experiments~\cite{dossantos2007}. 

In the present work, we explore how applied pressure tunes the optical response of Li$_{0.9}$Mo$_6$O$_{17}$. This allows us to assess the involved changes in the electronic conduction channels. Our most important finding is that the marked anisotropy of ambient-pressure optical properties vanishes above 6 GPa, concomitant to a structural phase transition. Overall, Li$_{0.9}$Mo$_6$O$_{17}$ undergoes several phase transitions under pressure confirmed by our x-ray diffraction data, further enriching its complex phase diagram. In the high-pressure phase, this compound appears to be no longer a quasi-1D conductor.

\begin{figure}[!!b]
\centering
\includegraphics[width=\columnwidth]{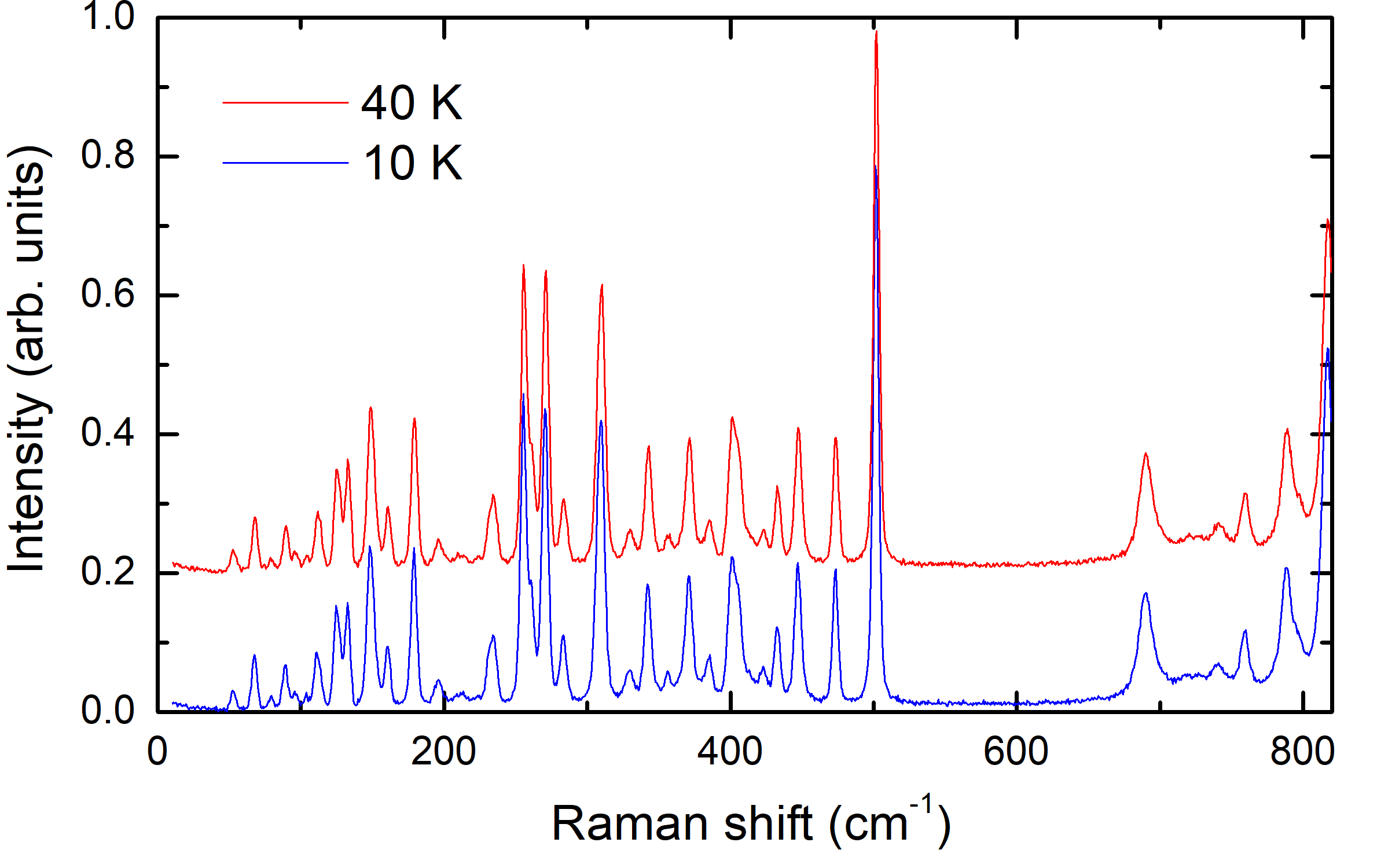}
\caption{Ambient-pressure unpolarized Raman spectra at 40~K (red curve) and 10~K (blue curve).} 
\label{figure3}
\end{figure}
\section{Results}
We measured x-ray diffraction, Raman and infrared spectra of Li$_{0.9}$Mo$_6$O$_{17}$ crystals as a function of pressure in the pressure range 0-15 GPa (Raman and infrared), and 0-10 GPa (x-rays).
The crystal growth procedure used for the  Li$_{0.9}$Mo$_6$O$_{17}$ crystals is described in Refs.~\onlinecite{greenblatt1988,mccarroll1984}.  
\begin{figure}[!!t]
\centering
\includegraphics[width=\columnwidth]{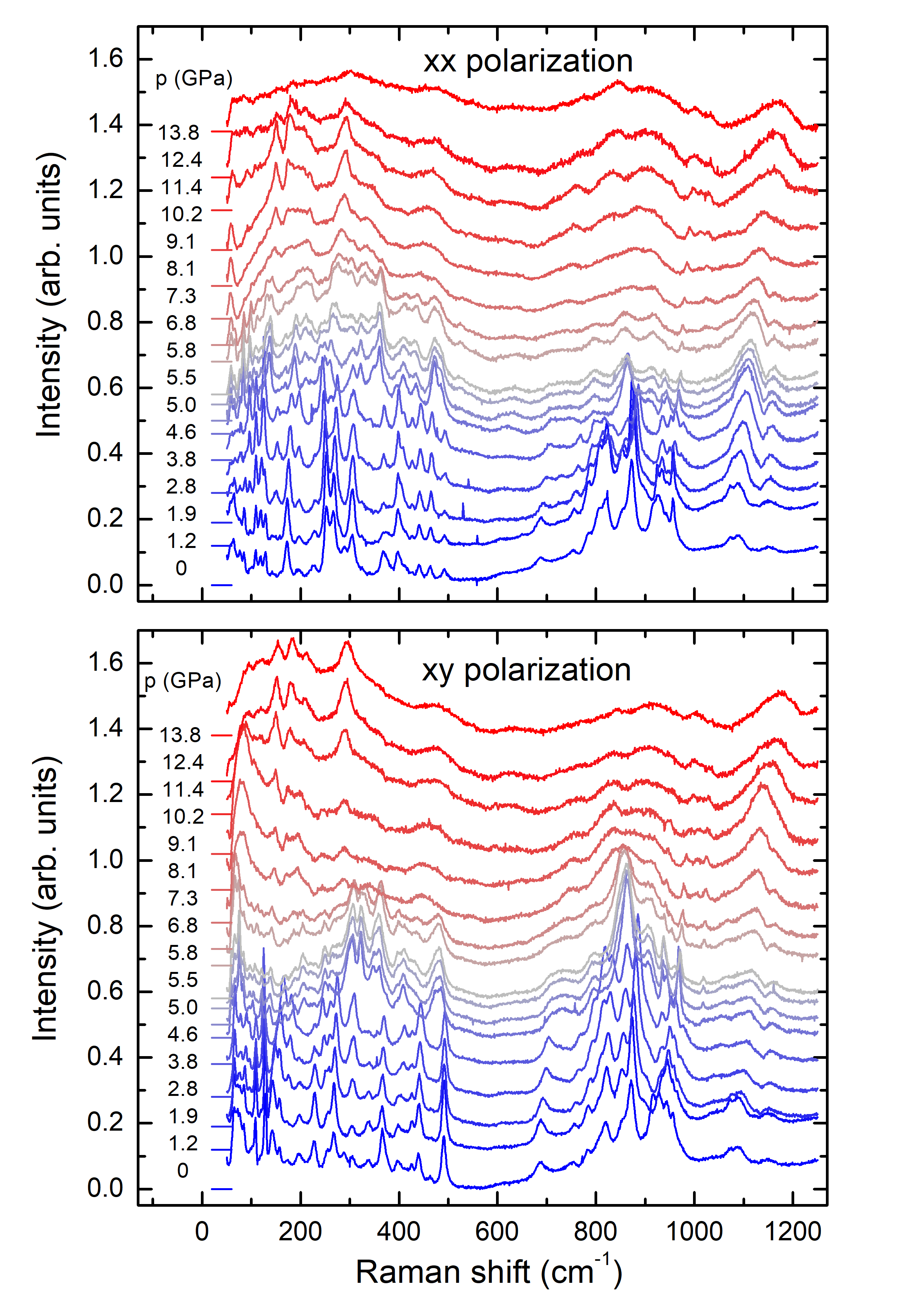}
\caption{Raman spectra for $XX$ (top)and $XY$ (bottom) polarization. The curves have been given vertical offsets proportional to the applied pressure. The baseline for each pressure is indicated as a horizontal line at the lefthand side of the spectra and pressures are indicated in the color coding of each spectrum.} 
\label{figure4}
\end{figure}
{\em X-ray diffraction.}
Single crystals of approximate dimensions $30\times40\times20\mu$m were first characterized using home-based x-ray diffractometers. Subsequently synchrotron x-ray diffraction was performed under pressure in diamond anvil cells both on single crystals as well as a powder ground from the same sample batch.
These measurements were done in two successive sessions at the High-Pressure Beamline ID~27 and at the Swiss Norwegian Beamline BM~01 of the European Synchrotron Radiation Facility ESRF.
The high-pressure experiments at ID~27 were performed at a wavelength of $0.3738~\AA$, using BETSA-type membrane diamond anvil cells (opening angle $68^\circ$) and He as a pressure medium. 
At BM~01, $\lambda = 0.68043~\AA$, samples were loaded into ETH-type diamond anvil cells~\cite{periotto2011,miletich2000} (opening angle $90^\circ$) and pressurized by manually tightening 4 screws, in an ethanol-methanol pressure medium, which has a hydrostatic pressure limit of $\sim10-11$~GPa. 
In all experiments the samples, crystals or powder were loaded into $600~\mu$m laser-drilled sample chamber, using pre-indented stainless steel gaskets of $\sim80~\mu$m thickness. 
Pressure was monitored in all cases with ruby spheres of $\sim3~\mu$m using the ruby fluorescence line. 
The powder diffraction experiment was used to detect phase transformations (see Appendix) as well as to solve the general topology of the phase between 3 and 6~GPa (data used: 5.9 GPa). 
The evolution of normalized lattice parameters is shown in Fig.~\ref{figure2}. 
\begin{figure}[!!t]
\centering
\includegraphics[width=\columnwidth]{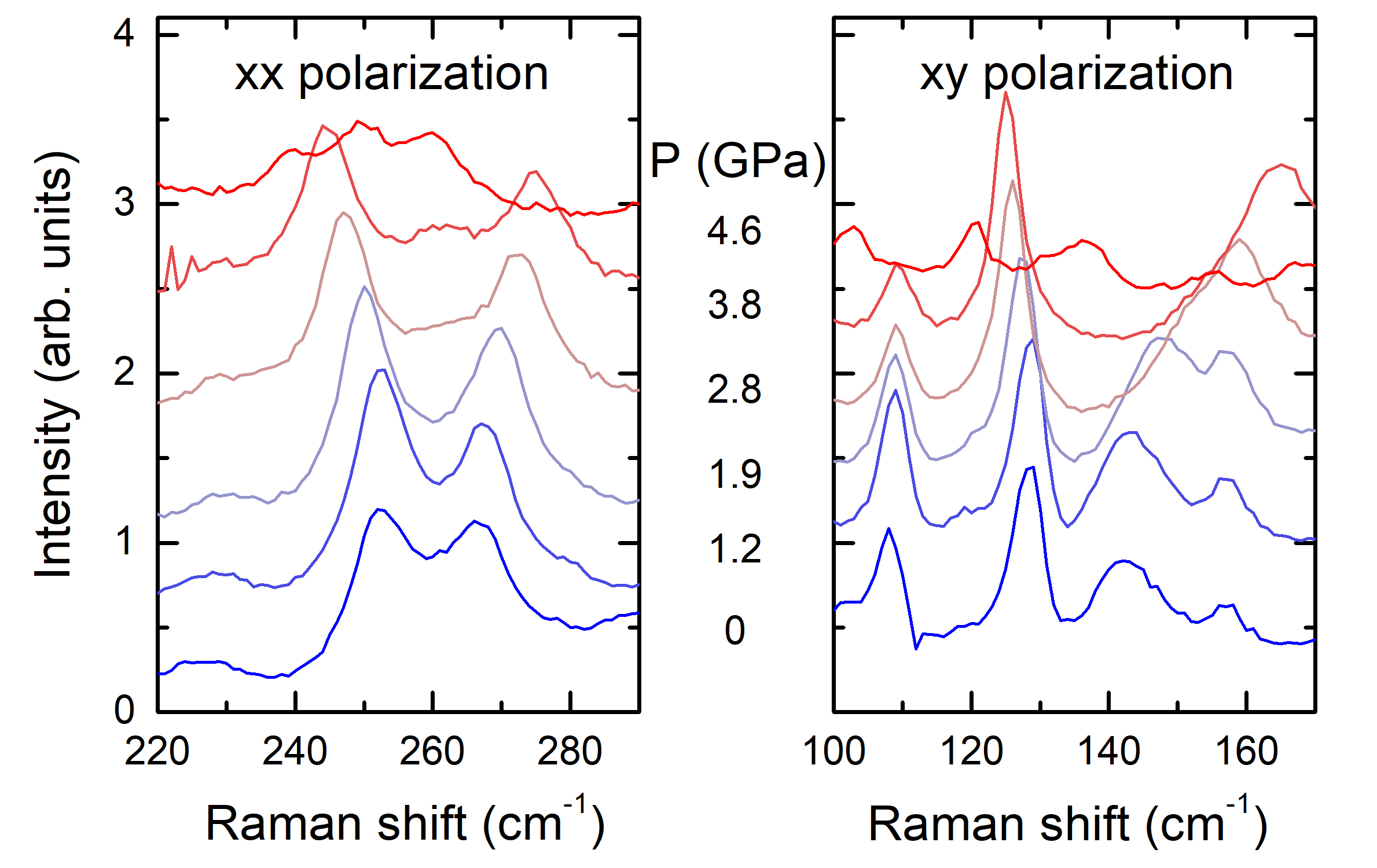}
\caption{Close-up of the Raman spectra for the XX (a) and XY (b) polarization. The curves have been given vertical offsets in proportion to the applied pressure.} 
\label{figure5}
\end{figure}
{\em Raman spectra} were measured on the $bc$ cleavage plane in backscattering geometry at room temperature as a function of pressure in the same pressure cell, using Daphne oil 7373 as a pressure transmitting medium and ruby chips as pressure sensors. 
The addition of a rotating half-wave plate allowed to disentangle the spectral contribution of $XX$ and $XY$ polarisations. 
It was, however, not possible to match the crystallographic $b$ and $c$ axes with the polarisation of the laser beam.
The Raman spectra at ambient pressure are displayed in Fig.~\ref{figure3}. 
The 10~K spectrum shows a large number of sharp phonon modes, consistent with a large unit cell and low symmetry of the compound. 
Panel (b) shows that the modes do not exhibit any noticeable changes with temperature, more specifically when crossing the temperature of the upturn in the resistivity. 
Due to the fact that a CDW breaks the translational symmetry of the crystal, a transition into a CDW state would be accompanied by the appearance of additional Raman modes. 
The absence of additional modes below 25~K  in the spectra of Fig.~\ref{figure3} is a strong indication that the resistivity upturn below 25~K is not caused by a CDW.
The evolution with pressure of the Raman spectra obtained for $XX$ (a) and $XY$ polarization (b) are presented in Figs.~\ref{figure4} and \ref{figure5}.

\begin{figure}
\centering
\includegraphics[width=\columnwidth]{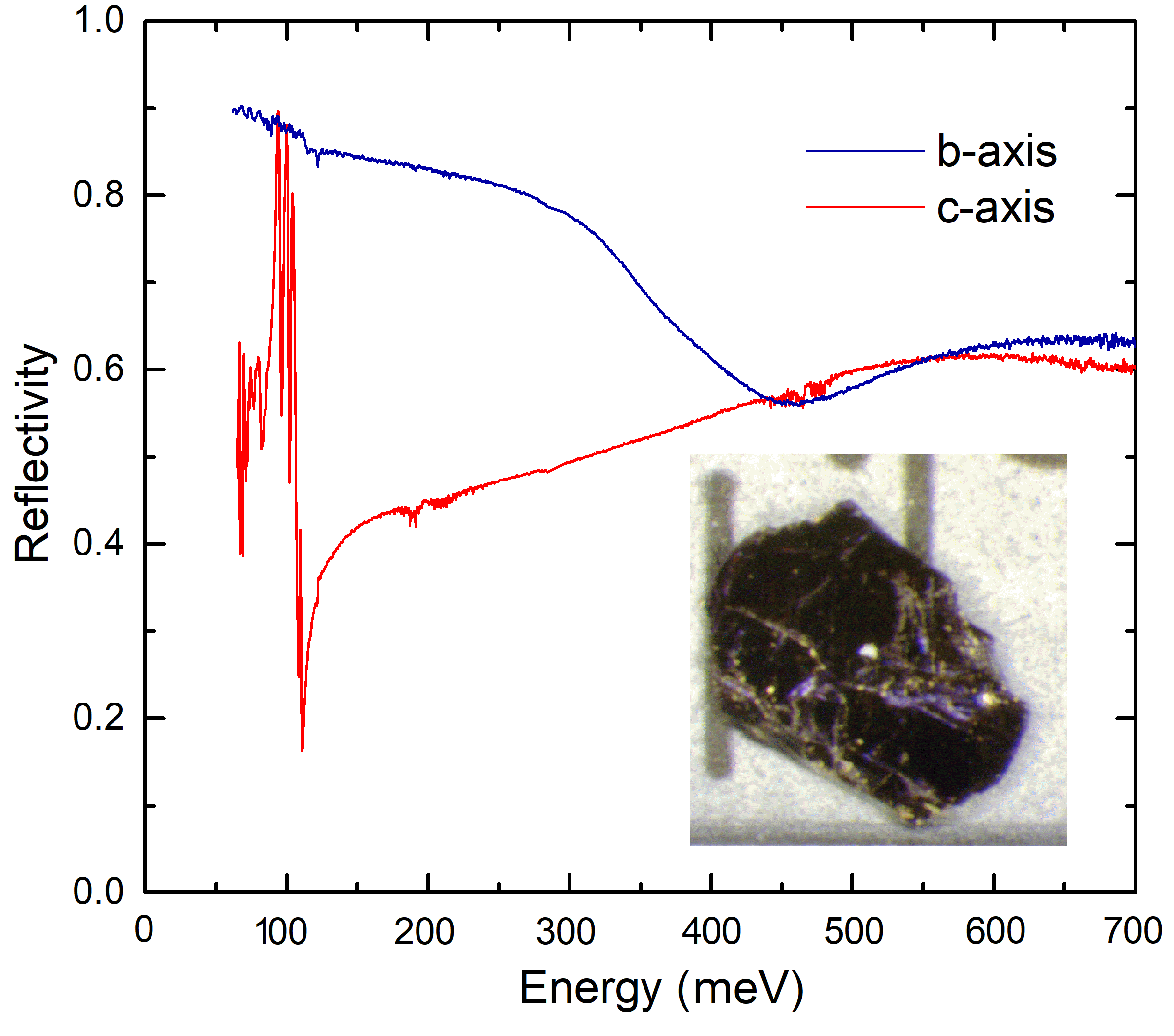}
\caption{Room-temperature reflectivity along the insulating and conducting axes at ambient pressure.  Inset: One of the crystals used in the present study.
\label{figure6}} 
\end{figure}

{\em Optical spectroscopy.}
Ambient pressure infrared reflectance spectra were obtained by measuring the intensity reflected from the sample at ambient pressure and calibrated against a gold layer evaporated {\em in situ} on the sample surface (Fig.~\ref{figure6}). 
Crystal surfaces were oriented parallel to the $bc$ plane and  a rotating gold wire polarizer was used to isolate the $b$ and $c$ axis components of the reflectivity tensor. 
The ambient pressure reflectivity at room temperature is shown in Fig.~\ref{figure6} for electric field polarized along the $b$ and $c$ axis respectively. 
This result reproduces the previous spectra determined by Choi {\em et. al}~\cite{choi2004}. 
A strong anisotropy is evident from the reflectivity data, where the $b$ axis is metallic, and the $c$ axis insulating.
The $b$ axis shows reflectivity that tends to unity as photon energy decreases, consistent with a metallic response. A broad shoulder from 300 to 500~meV may be linked to the screened plasma edge. In contrast, the $c$ axis shows sharp phonon features that dominate the low-energy reflectivity. The overall low-energy value of the reflectivity is significantly smaller than along $b$ axis, pointing to an insulating response.

\begin{figure}
\centering
\includegraphics[width=\columnwidth]{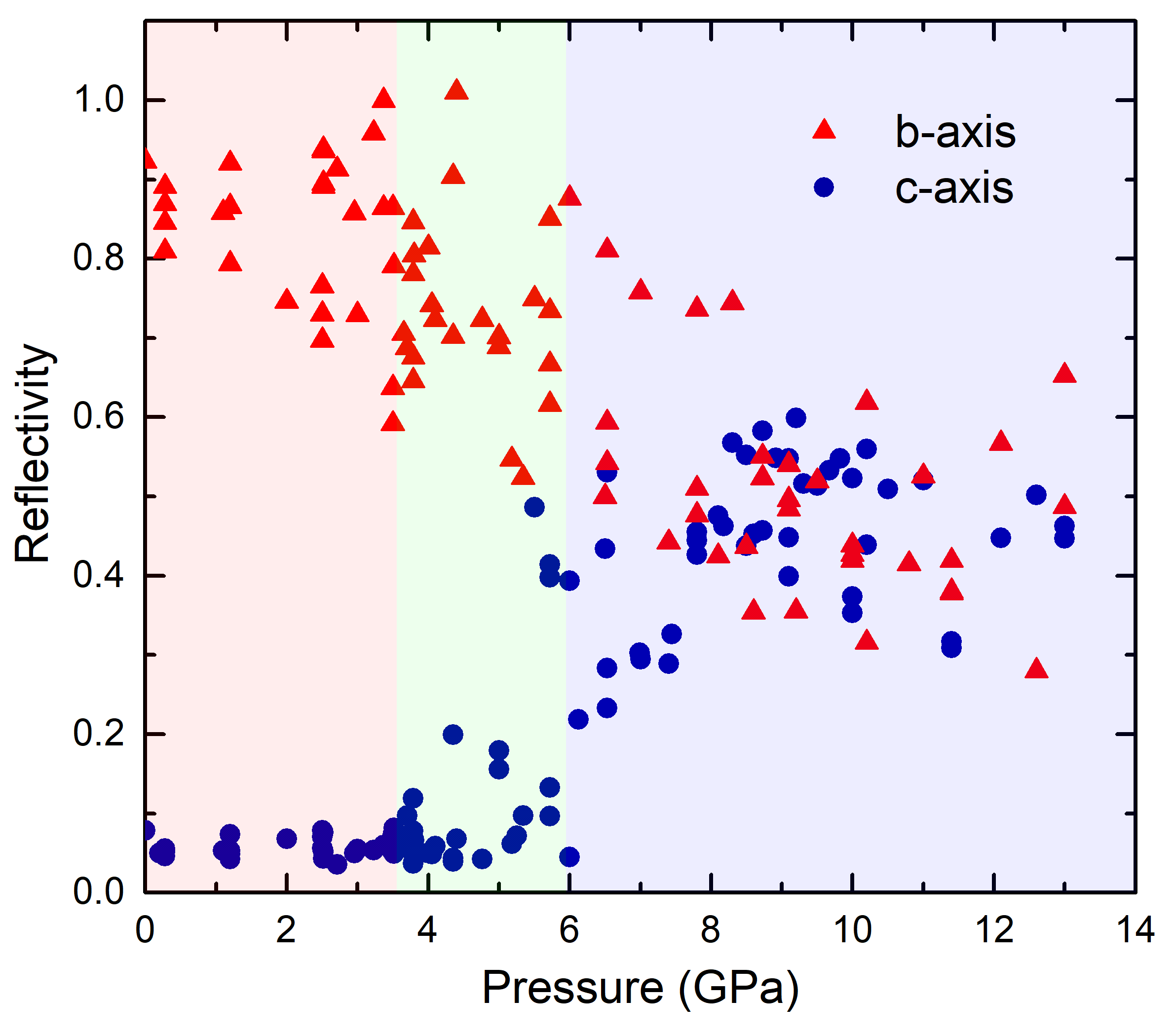}
\caption{Pressure dependence of the reflectivity at 124 meV (1000 cm$^{-1}$) for $b$ (red triangles) and $c$ (blue dots) axis.
\label{figure7}}
\end{figure}

High pressure infrared reflectivity spectra were acquired at the infrared beamline of the Swiss Light Source~\cite{lerch2012} using a diamond anvil cell-based customized setup~\cite{tran2015} fitting a Hyperion microscope. 
Freshly cut~$\sim0.03\times0.04\times0.01$~mm samples with the surface oriented parallel to the $c$ and $b$ axes were loaded in the sample chamber consisting of CuBe gaskets with ruby chips. We used KBr powder as a pressure transmitting medium. 
Since with this procedure the samples are pressed against the diamond window, the measured reflectivity spectra are those of a sample/diamond interface.
A limited spectral range of 50-600~meV is imposed by the presence of a KBr window between the spectrometer and the microscope as well as the use of a gold wire FIR-MIR polarizer located at the rear of the reflective objective. 
As the pressure was tuned at ambient temperature, a full thermal cycle is needed before changing the pressure inside the diamond anvil cell. This, together with the mechanical stabilisation time, reduces the number of accessible pressure points at low temperature. 
To demonstrate the effect of pressure we begin with the reflectivity at 124~meV as a function of pressure, shown in Figure~\ref{figure7} for light polarized both along $b$ and $c$ axis. It should be mentioned at this point that, due to the departure from $90^\circ$ of the angle between $c$ and $a$ the optical response of these two axis is strictly speaking not separated. The reflectivity polarized along $c$ provides a pseudo-dielectric function of mixed $ac$ character, and to correct for the mixing requires a combination of spectra on different crystal faces.  In the present case the angle of $90.6^\circ$ is sufficiently close to $90.0^\circ$ to render these corrections too small to be relevant in the context of the present study. 
The photon energy is selected such that it is higher than the observed phonon modes, and a large difference can be seen in $b$ and $c$ polarizations.
The reflectivity along the initially more metallic $b$ axis drops continuously. Consistent with the notion that metals make good mirrors due to the dielectric properties of free charge carriers, this drop of reflectivity indicates that the $b$-axis becomes less conducting under pressure.
On the other hand, the reflectivity along the initially insulating $c$ axis becomes progressively higher above $\sim 4$~GPa, and finally reaches similar values as the $b$ axis reflectivity above 8~GPa. In principle one should expect that the reflectivity progressively approaches the 100 \% level for $\omega \rightarrow 0$.  
Whether this behavior results from a transition toward a more conducting state along the $c$-axis and a less conducting one along $b$, requires an analyses of the broad reflectivity spectra that we will now present. 

\begin{figure}[!!t]
\centering
\includegraphics[width=\columnwidth]{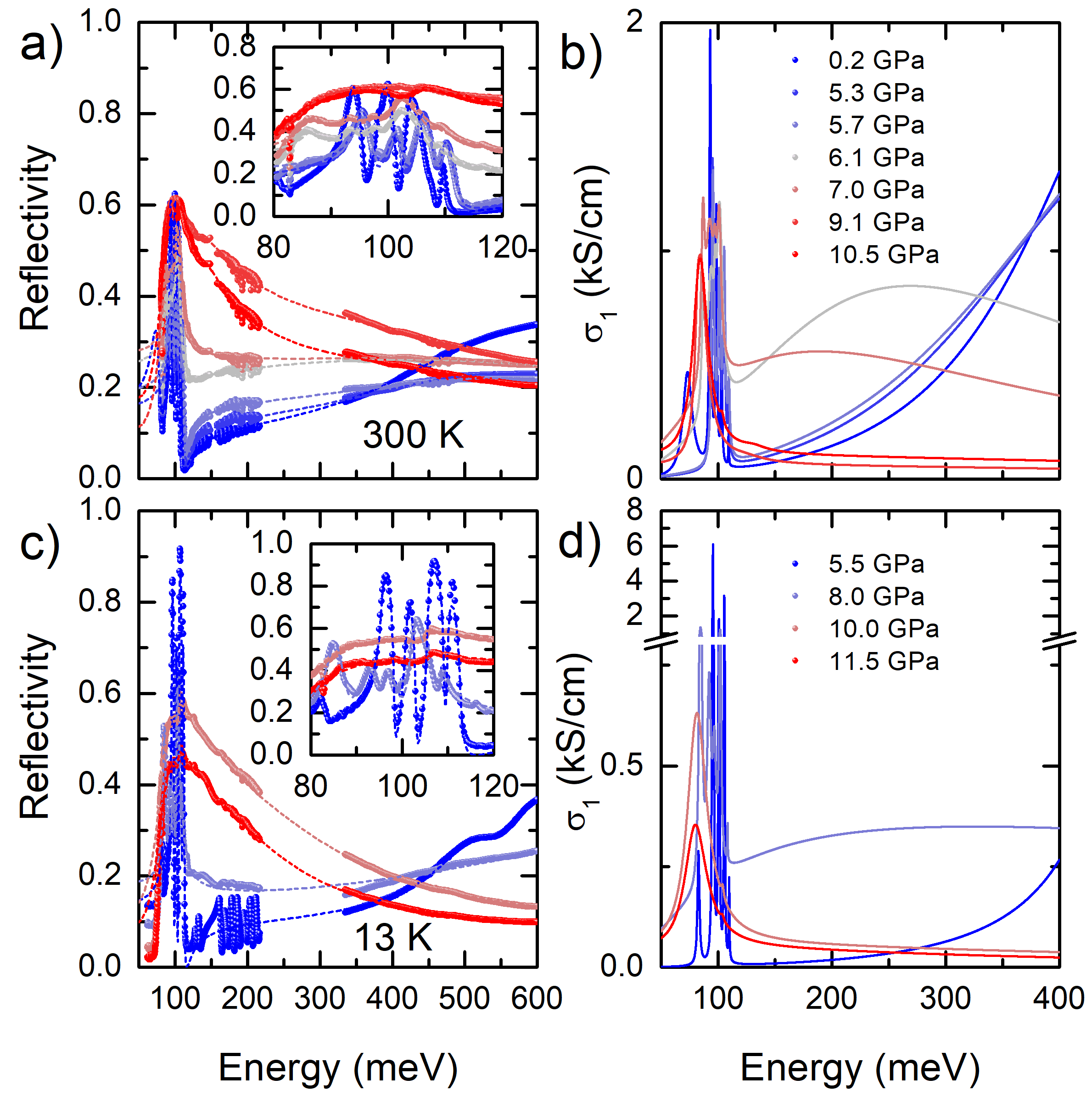}
\caption{Pressure-dependent evolution of the $c$-axis reflectivity and optical conductivity at room temperature and at 13~K. 
\label{figure8}}
\end{figure}

\begin{figure}[!!t]
\centering
\includegraphics[width=\columnwidth]{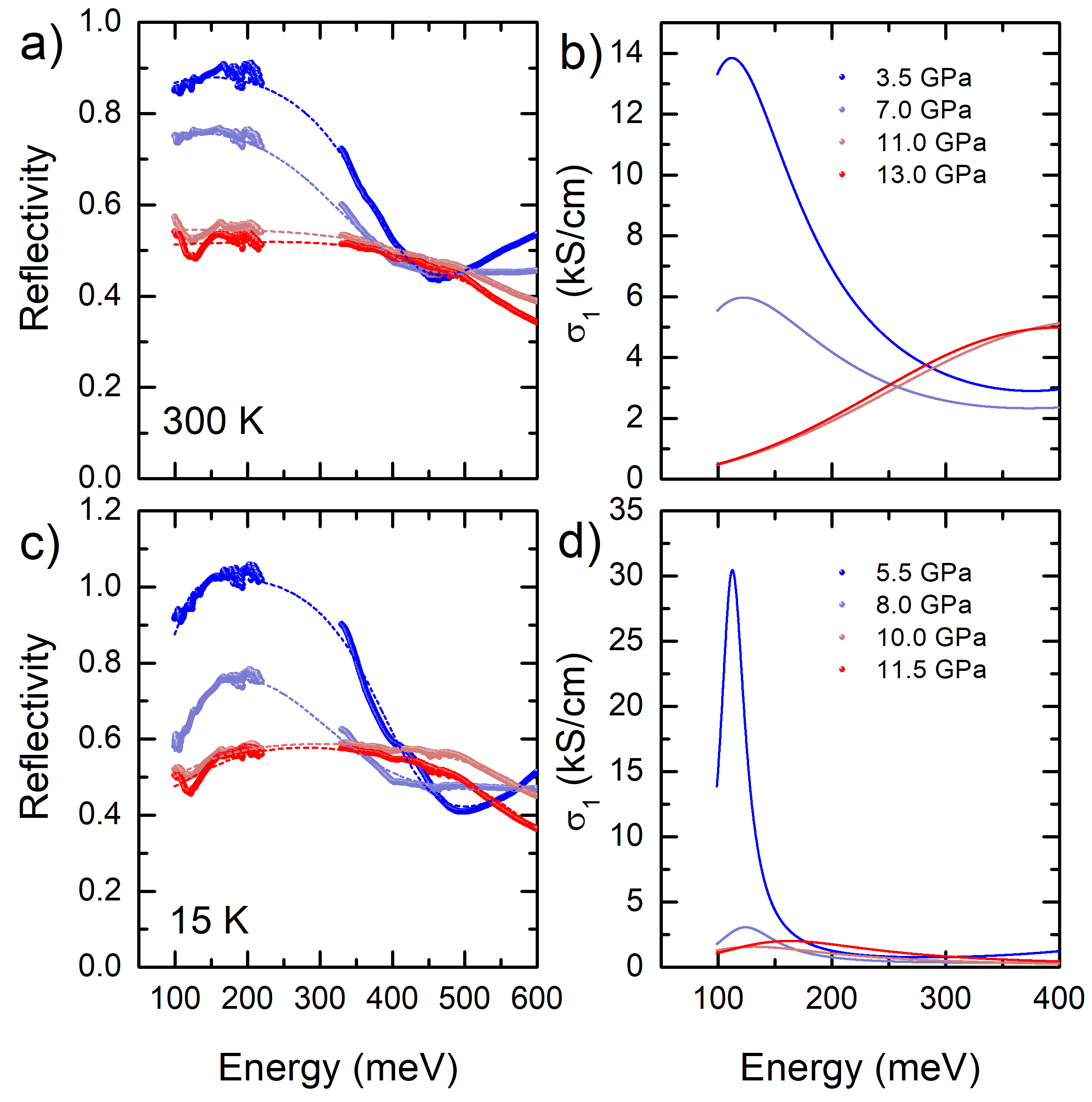}
\caption{Pressure-dependent evolution of the $b$-axis reflectivity and optical conductivity at 300~K and at 15~K.  
\label{figure9}} 
\end{figure}

The reflectivity spectra for selected pressures with electric field polarized along along the $c$ axis and $b$ axis are displayed in Figs.~\ref{figure8} and ~\ref{figure9} respectively. We calculated the optical conductivity  $\sigma(\omega)=i\omega[1-\epsilon(\omega)]/4\pi$ by fitting the reflectivity $R(\omega)$ 
\begin{equation}
   R(\omega)=\left|\frac{n_d-\sqrt{\epsilon(\omega)}}{n_d+\sqrt{\epsilon(\omega)}}\right|^2
  \label{eq1}
\end{equation}
where $n_d\approx 2.4$ is the optical constant of the diamond window, to the Drude-Lorentz expansion
\begin{equation}
    \epsilon(\omega)=1-\sum_j\frac{\omega_{p,j}^2}{i\omega\gamma_j+\omega^2-\omega_j^2} 
      \label{eq2}
\end{equation}
At the lowest pressures, these data are qualitatively consistent with previous ambient pressure studies~\cite{choi2004}. 
For all pressures from 0 up to 5.5~GPa the $c$-axis reflectivity shown in Fig.~\ref{figure8} is characteristic of an insulator, with strong phonon modes and a low background level. 
At the highest pressures two phonon modes are evident, superimposed on a higher background.
The main effect of cooling to 13~K (see Fig.~\ref{figure8}, panels (a) and (c)) is a sharpening of the phonons and small shifts of the frequencies. 
From the optical conductivity shown in panels (b) and (d) we see that the intensities of the phonons do not change much as a function of pressure. More significant is the presence in the high pressure phase of continuum of excitations from the lower end of the measured frequency range (80~meV) and possibly lower, to at least 200~meV or higher (note that due to optical absorption of the diamond, the reflectivity spectrum is cut between 200 and 300~eV). This continuum is clearly electronic in nature and could either be due to transitions across a small semiconductor gap (80~meV or smaller) or intraband conductivity of strongly interacting electrons~\cite{basov2011}. Either interpretation indicates a significant difference of the electronic properties along the $c$-axis between low pressure (Fig.~\ref{figure1}a,c) and high pressure (Fig.~\ref{figure1}b,d), namely a change from insulating behavior in the $P2_1/m$ phase to a small gap semiconductor or a bad metal in the $P2_1/a$ phase. 

Contrary to the $c$-axis, the reflectivity of the $b$-axis displays no strong phonon-like spectral features. At room temperature, the $\sigma_1(\omega)$ data seems to imply a gradual decrease in the Drude contribution. At low temperatures, the Drude contribution disappears entirely from our pressure and photon energy window.
Under 5.5~GPa, a prominent feature appears -- a narrow band at approximately 100~meV. The spectral weight exceeds typical phonon values by 3 orders of magnitude. The character of this band is therefore mainly electronic. We interpret this as a side-band of the Drude peak resulting from electron-phonon or electron-electron interactions~\cite{basov2011}. 

\section{Discussion}
The x-ray diffraction, Raman and infrared optical data reveal that as a function of pressure the system passes through a succession of different phases. These phases are summarized in Table \ref{table1} along with the crystallographic data obtained in the present study.

\begin{figure}
\begin{center}
\includegraphics[width=\columnwidth]{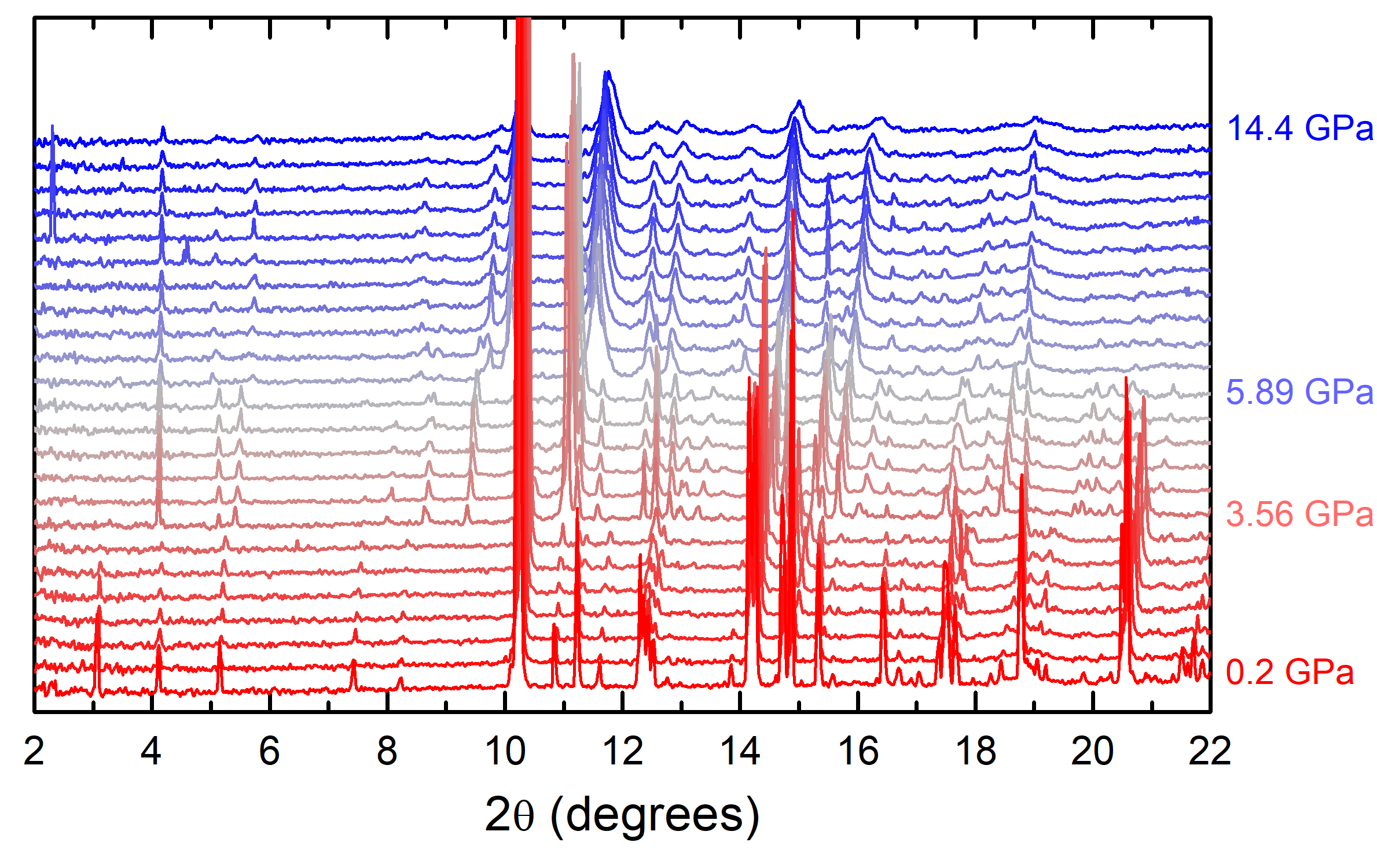}
\caption{Pressure ramp collected on powdered sample at BM~01, $\lambda = 6.8043~\AA$, showing 2 phase transitions.}
\label{figure10}
\end{center}
\end{figure}

{\em Between ambient pressure and 3.0~GPa.} only the $P2_1/m$ phase is observed. 
The polyhedral drawing of the structure refined at $1.27$~GPa is shown in Fig.~\ref{figure1}. 
It appears that the buckling of corner-sharing octahedral chains in the ambient pressure phases increases upon pressure application (increase of Mo-O-Mo angle between neighboring octahedra). 
Powder diffraction shows a structural transition most readily seen at scattering angles around $2\theta=5.2^\circ$, and manifested by a pronounced peak splitting at 3.6~GPa (Figs.~\ref{figure2} and~\ref{figure10}).  
In the ambient pressure phase, this intensity is composed of the $101$ and $10\overline{1}$ Bragg reflections, which mark the monoclinic distortion and overlap entirely (within instrumental resolution) owed to the minor monoclinic angle of $90.66^\circ$. 
At the transition, this angle increases significantly to $93.5^\circ$, accompanied by a structural distortion and rearrangement of coordination polyhedral, which is at the origin of the volume collapse (Fig.~\ref{figure2}) of $\sim8\%$ at this first transition. 
This rather large increase is surely triggered by the relative compliance of the coordination polyhedron formed around the Li atom\footnote{Li is omitted from the structural drawings as it could be located neither from powder nor from single crystal diffraction under pressure}. 

{\em Between 3.0 and 3.6~GPa} $P2_1/m$ and $P2_1/a$ phases coexist as seen in the Rietveld plot at 3.6~GPa (see Appendix).  The phase transition in this pressure range strongly affects the Raman spectrum. In particular two groups of modes, one between 100 and 200 cm$^{-1}$ in XY polarization and the other between 900 and 900 cm$^{-1}$ in both polarizations, are present below the critical pressure and disappear completely above 4~GPa (see Figs.~\ref{figure4} and~\ref{figure5}). 
In the infrared reflectivity along the $c$-axis (Fig.~\ref{figure8}) additional phonon features appear for pressure in this range and the overall appearance of the phonon spectrum changes. 
At 300~K the phonon spectrum becomes drastically rearranged between 0.2 and 6.1~GPa. At 13 ~K the rearrangement happens between 5.5 and 8~GPa, indicating that at low temperatures the structure is more resilient against pressure induced changes than at 300~K.
This change happens rather gradually, which is consistent with the coexistence of two phases. 

{\em Between 3.6 and 6.0~GPa} (HP1) only the $P2_1/a$ phase is observed. This space group is confirmed by single crystal diffraction data in the reciprocal space reconstructions of the $0kl$, $1kl$ and $-1kl$ planes at 4.14~GPa (see Appendix). The primitive cell of the $P2_1/a$ structure contains 4 formula units as compared to 2 formula units for the $P2_1/m$ structure at low pressure. Correspondingly the $c$-axis of the $P2_1/a$ structure is doubled as compared to $P2_1/m$. Lattice parameters obtained from LeBail fits to powder data are given in Table~\ref{table1}. 
During this symmetry change all Mo coordination centers move away from the special Wyckoff site $2e$ on the mirror plane at $y = 1/4$ in the ambient pressure phase $(P2_1/m)$ to the lower symmetry general position $4e$ in the $(P2_1/a)$-phase. 
At the transition pressure, one tetrahedral and one octahedral position merge into edge-sharing dimers, which form edge-sharing chains along the $b$-axis in the $P2_1/a$-phase (light-green in Fig.~\ref{figure1}). 
A strong buckling of the surrounding corner-sharing octahedral zig-zag chains is observed. 
It is known that not only changes in connectivity as the present ones, but also changes in octahedral tilting affect the band dispersion within the band structure.  
The most significant structural change for the transport properties is the rearrangement of the MoO$_4$ tetrahedra and part of MoO$_6$ octahedra in the $P2_1/a$-phase, due to which MoO$_6$ octahedra in adjacent slabs now share corners connecting the slabs along the $a$ direction. The band structure of this new phase has not yet been reported. 
The aforementioned rearrangement of the MoO$_4$ tetrahedra, in particular the corner sharing along $a$, increases the hopping between the slabs, and introduces hopping paths of MoO$_6$ octahedra along the $c$-axis which are absent in the $P2_1/m$-phase. 
All in all this is expected to make the band dispersion in the $P2_1/a$-phase more 3-dimensional than in the $P2_1/m$-phase, which is qualitatively consistent with our optical data.

{\em Above 6~GPa } (HP2) a new crystallographic phase is formed (Fig.~\ref{figure2}) whose primitive lattice, closely related to the one of the $P2_1/a$ phase, we were able to determine (see Table~\ref{table1}) but whose atomic structure we could not solve. Lattice parameters were nevertheless refined from the powder diffraction data with a LeBail fit.
Indexation of the data (not high quality) no longer shows evidence of the superstructure in $c$.
This structural transition appears to be triggering the dimensional crossover. 
Interestingly, an analysis of reciprocal space reconstructions reveals 
a doubling of the $b$-axis possibly due to a CDW (see Appendix, weak intensities at $k = -1.5$), for which pressures we observe a gradual decrease of conductivity. 
For all pressures from 0 up to 5.5~GPa the $c$-axis optical reflectivity spectra (Fig.~\ref{figure8})  display the characteristic features of optical phonons in the insulating state, whereas at higher pressure most phonon features disappear while the reflectivity rises with increasing pressure. The reflectivity spectrum above 5.5~GPa is characteristic of a metal, with a fairly high value and weak phonon peaks, where the dielectric function is dominated by the electronic screening.
The pressure induces significant spectral changes in the $b$-axis optical conductivity (Fig.~\ref{figure9}). 
Above 5.5~GPa the $b$-axis reflectivity strongly decreases, while the band in $\sigma_1(\omega)$ at 100~meV broadens and its intensity gets suppressed at approximately 7~GPa. Taken together this behavior of the optical response is indicative of a loss of free carrier spectral weight along the $b$-axis in the high pressure phase above 5.5~GPa.   

The positive slope of the conductivity at 400~meV visible in Figs.~\ref{figure8} and ~\ref{figure9} for pressures below 6~GPa gets suppressed at higher pressures. Instead, a broad band appears in the range below 400~meV.
This is indicative of a rearrangement of the electronic bands. This has the effect that interband transitions seen along the $b$ axis at 500~meV disappear, the material becomes a less good conductor along $b$ and a better conductor along $c$ axis. All in all this implies that the phase above 5.5~GPa is electronically more isotropic with a metallic conductivity along the $b$ and $c$ axis.

{\em 8.5~GPa} is the highest pressure for which single crystal x-ray diffraction data could be collected. Here the $hk\overline{1}$ plane shows splitting of Bragg reflection related to structural changes in the $ab$-plane (see Appendix), which can at present not be investigated due to the lack of data quality. 
It should be noted that all structural changes observed up to the highest pressures are fully reversible on single crystal samples, and refinements collected outside the diamond anvil cell in previously pressurized crystal yield good $R_{1}$ values of approx. 3\%, which are comparable to those collected prior to pressurization and demonstrate a very high reversible flexibility of the lattice.
The Raman spectra above 9~GPa (Fig.~\ref{figure4}) are characterized by a strong broadening and loss of intensity of the modes in XY polarization.

\section{Conclusions}
The application of high pressure on Li$_{0.9}$Mo$_6$O$_{17}$ strongly affects its crystal structure which remains in its ambient $P2_1/m$ phase up to near 3~GPa where it  changes to the $P2_1/a$. This phase  lasts up to 6~GPa. At 6~GPa a transition takes place to yet another monoclinic phase which is metallic both along $b$ and $c$ axis. 
This trend is observed at room temperature and at 13~K. This behavior is not surprising, since pressure generally tends to increase the orbital overlap favoring increased mobility, as observed in many quasi-1D organic compounds. An intriguing and counterintuitive aspect is the observation of a decreasing conductivity along the conducting axis. This apparent pressure-induced decrease in anisotropy is a clear indication that Li$_{0.9}$Mo$_6$O$_{17}$ is driven away from its initial quasi-1D dimensional conduction scenario. 

The datasets generated and analyzed during the current study are available in Ref.~\onlinecite{yareta2021}. These will be preserved for 10 years. All other data that support the plots within this paper and other findings of this study are available from the corresponding author upon reasonable request.

\section{Acknowledgments}
We thank Thierry Giamarchi and Radovan {\v C}ern{\'y} for stimulating discussions. JWA thanks O. K. Andersen for illuminating discussions of the two $a-c$ axis labeling schemes in the literature. We are grateful to Medhi Brandt for technical assistance.
This project was supported by the Swiss National Science Foundation through projects 200020-135085, P00P2$\_$170544 and 200020-179157.
We are grateful for beam time allocation by ESRF and SNBL, and acknowledge local support by Paraskevas Parisiades from ID 27 and Vladimir Dmitriev from SNBL.

\begin{figure}[!!t]
\begin{center}
\includegraphics[width=\columnwidth]{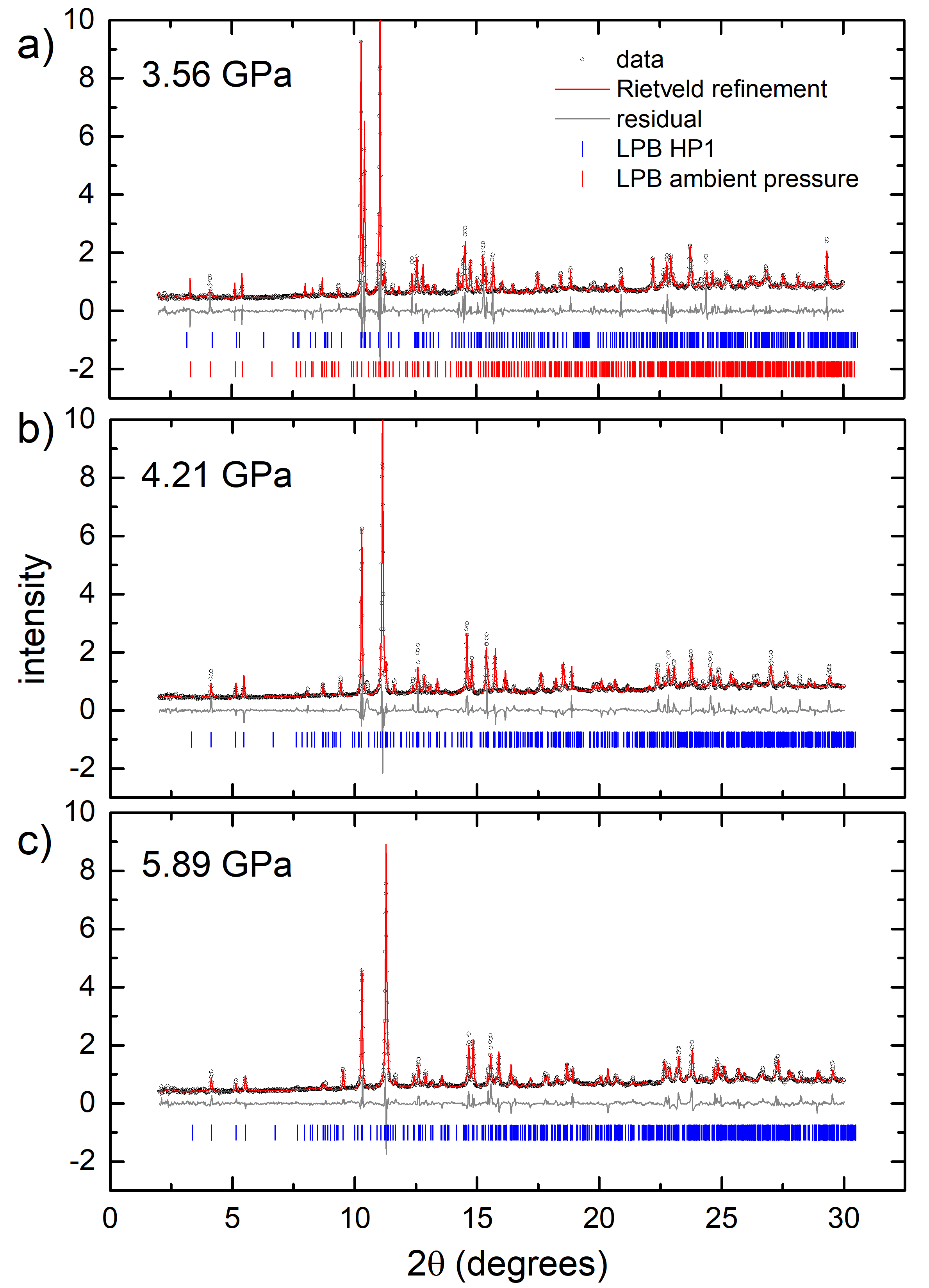}
\caption{Rietveld refinement at a) 3.56~GPa, b) 4.21~GPa and c) 5.89~GPa (bottom), with $\lambda=0.68043~\AA $. Symbols are Li$_{0.9}$Mo$_6$O$_{17}$ diffraction data, red curves Rietveld refinement and grey curves are fit residuals.  Vertical line represents the positions of diffraction pics of Li$_{0.9}$Mo$_6$O$_{17}$ in the ambient pressure structure (red) and in the high pressure HP1 structure (blue).}
\label{figure11}
\end{center}
\end{figure}
\appendix
\section{X-ray diffraction}\label{supXRD}
 The powder diffraction data were integrated to 1D diffraction patterns using Fit2D~\cite{hammersley1994}. An example for different pressures is shown in Fig.~\ref{figure11}). Direct Space modeling was done in FOX~\cite{cerny2007}, data were then refined with the Rietveld method using Topas5~\cite{coelho2018}. During Rietveld refinement anti-bump restraints were applied to the Mo-Mo and Mo-O interatomic distances. No texture corrections were required, evidencing hydrostatic pressure conditions in the pressure chamber. We note that the Li position occupied to 90\% was omitted from refinements, as it was not visible by x-ray diffraction and/or its positional shifts did not behave in a stable way during minimization. This is not a surprise given the inherently lower quality of high-pressure diffraction data and the lack of contrast between the relatively light Li atom and its surroundings. Selected Rietveld refinements of the first high-pressure phase (3-6~GPa) are shown in Fig.~\ref{figure11}.

\begin{figure*}
\begin{center}
\includegraphics[width=\textwidth]{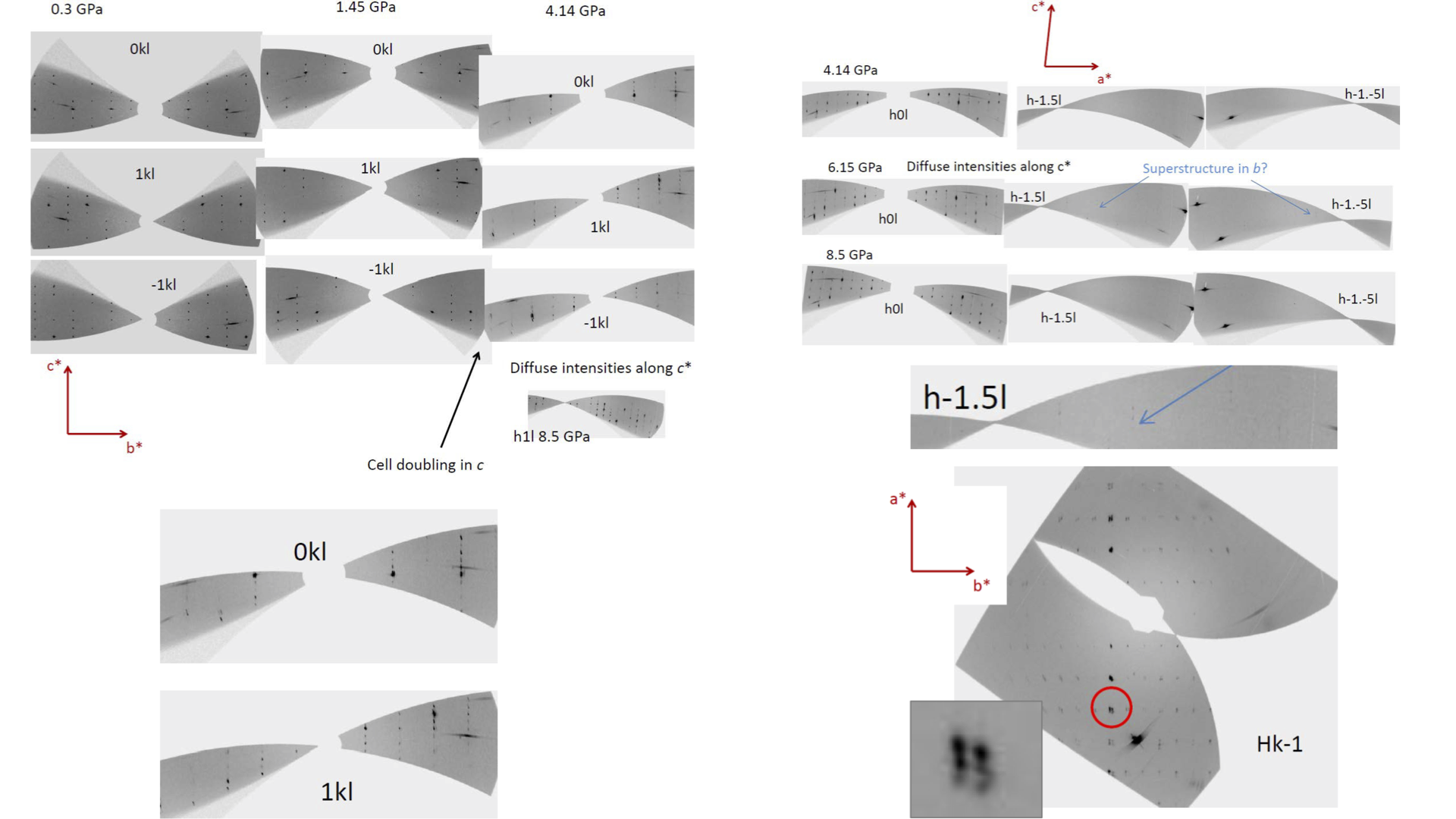}
\caption{Reciprocal space reconstruction for different pressures. For all panels $\lambda = 0.3738~\AA$. The data show cell doubling of $c$ at 4.14~GPa, doubling of $b$ at 6.15~GPa and Bragg intensity splitting in the $a-b$-plane at 8.5~GPa (lower right image). }
\label{figure12}
\end{center}
\end{figure*}

Single crystal data were treated with the SNBL tool~\cite{dyadkin2016} and CrysAlis(Pro)~\cite{crysalispro} and refined using shelxl~\cite{sheldrick2015} through the Olex2~\cite{dolomanov2009} user interface. Reciprocal space reconstructions (Fig.~\ref{figure12}) were done with CrysAlis(Pro).
Single crystal refinements were performed on BM~01 data only, as the collected number of reflections was higher, due to the larger opening angle of the diamond anvil cell and the more favorable orientation of the loaded crystal. Data integration on the respective data sets yielded $R_{int}$'s of 3-3.5\%, least squares refinement resulted in $R_{1}$ values of 11-12\%, which is acceptable for high-pressure experiments and a discussion of the general topology of the structure. The structure the phase between 3 and 6~GPa could not be solved directly from single crystal data, and was instead modeled by global optimization using FOX~\cite{cerny2007} on powder data obtained at 5.9~GPa. Despite the applied pressure, there is no evidence of preferred orientation, which supports the accuracy (at low resolution) of the model obtained by Direct Space methods. This obtained minimized model was then used as input model for Rietveld refinements as well as single crystal structure refinements.
\onecolumngrid
\twocolumngrid
%
%
%
%
%
\end{document}